\documentclass[twoside]{article}
\usepackage{fleqn,espcrc2}
\usepackage{graphicx}
\usepackage{epsfig}

\title{The Solar Neutrino Problem  in the Light of  a Violation  
of the Equivalence Principle\thanks{Talk given by R. Zukanovich Funchal.}}

\author{A. M. Gago\address{Instituto de F\'{\i}sica, Universidade de 
S\~ao Paulo, C. P. 66.318, 05315-970 S\~ao Paulo, Brazil}%
\thanks{On leave of absence from : Secci\'{o}n F\'{\i}sica, Departamento de Ciencias, Pontificia Universidad Cat\'olica del Per\'u,
Apartado 1761, Lima, Per\'u.}
H. Nunokawa\address{Instituto de F\'{\i}sica Gleb Wataghin, 
    Universidade Estadual de Campinas, 13083-970 Campinas, Brazil}
and 
R. Zukanovich Funchal$^{\mbox{\scriptsize a}}$}

\begin{document}

\begin{abstract}
We have found that long-wavelength neutrino oscillations induced 
by a tiny breakdown of the weak equivalence principle of general 
relativity can provide a viable solution to the solar neutrino problem.
\end{abstract}
            
\maketitle

\section{INTRODUCTION}

Neutrinos have had, since their childhood in the early 30's, profound
consequences on our understanding of the forces of nature. In the past they
led to the discovery of neutral currents and provided the first indication
in favour of the standard model of electroweak interaction. They may be today
at the very hart of yet another breakthrough in our perceptions of the
physical world.   

Today the results coming from solar neutrino 
experiments~\cite{homestake,sage,gallex,sk99} as well as 
from atmospheric neutrino experiments~\cite{atmospheric} 
are difficult to be understood without admitting neutrino flavour
conversion. Nevertheless the dynamics underlying such conversion 
is yet to be established and in particular does not have to be a priori 
related to the electroweak force. 

The interesting idea that gravitational forces may induce neutrino mixing 
and flavour oscillations, if the weak equivalence principle
of general relativity is violated, was proposed  
about a decade ago~\cite{gasper,hl}, 
and thereafter, many works have been performed on this
subject~\cite{muitas}. 

Many authors have investigated the possibility of solving the solar
neutrino problem (SNP) by such gravitationally induced  neutrino 
oscillations~\cite{pantaleone,bkl,kuo}, generally finding it necessary,
in this context, to invoke the MSW like resonance~\cite{hl} since they   
conclude that it is impossible that 
this type of long-wavelength vacuum
oscillation could explain the specific energy dependence 
of the data~\cite{pantaleone,bkl}.  Nevertheless we
demonstrate that all the recent solar neutrino data
coming from gallium, chlorine and water Cherenkov detectors
can be well accounted for by  long-wavelength 
neutrino oscillations induced by a violation of the equivalence 
principle (VEP).

\section{THE VEP FORMALISM}

We assume that neutrinos of different types will suffer
different time delay due to the  weak, static gravitational field in the
space on their way from the Sun to the Earth. 
Their motion in this gravitational field can be appropriately
described by the parameterized post-Newtonian formalism
with a different parameter for each neutrino type. 
Neutrinos that are weak interaction eigenstates and
neutrinos that are gravity eigenstates will be related by a unitary
transformation that can be parameterized, assuming only two neutrino
flavours,  by a single parameter, the mixing 
angle $\theta_G$ which can lead to flavour oscillation~\cite{gasper}. 

In this work we assume oscillations only between 
two species of neutrinos, which are degenerate in mass, 
either between active and active ($\nu_e \leftrightarrow
\nu_\mu,\nu_\tau$) or  active and sterile ($\nu_e \leftrightarrow \nu_s$, $\nu_s$ being an electroweak singlet) neutrinos. 

The evolution equation for neutrino flavours $\alpha$ and $\beta$ propagating 
through the  gravitational potential $\phi(r)$ in the absence of
matter can be found in Ref.~\cite{us}.
In the case we take $\phi$ to be a constant, this can  be analytically 
solved to give the survival probability of $ \nu_e$ produced in the Sun 
after travelling the distance $L$ to the Earth

\begin{equation}
 P( \nu_e \rightarrow \nu_e) 
= 1 - \sin^2 2\theta_G \sin^2 \frac{\pi L}{\lambda},
\label{prob}
\end{equation}
where the oscillation wavelength $\lambda$ for a neutrino with 
energy $E$ is given by
\begin{equation}
 \lambda 
= \left[\frac{\pi {\mbox{ km}}}{5.07}\right] \left[\frac{10^{-15}}
{|\phi \Delta \gamma|}\right] \left[\frac{ {\mbox{MeV}}}{E}\right],
\label{wavelength}
\end{equation}
which in contrast to the wavelength for mass  induced neutrino 
oscillations in vacuum, is inversely proportional to the neutrino energy.
Here $\Delta \gamma$ is the quantity which measures the magnitude 
of VEP.

\section{ANALYSIS}

We will discuss here our analysis and results for active to active conversion.
The same analysis for the $\nu_e \to \nu_s$ channel
can be found in Ref.\ \cite{us},  given similar results.

In order to examine the observed solar neutrino rates in the
VEP framework we have calculated the theoretical  
predictions for gallium, chlorine and Super-Kamiokande (SK) water Cherenkov  
solar neutrino experiments, as a function of the two VEP parameters ($\sin^2 2 \theta_G$ and $ | \phi \Delta \gamma |$), 
using the solar neutrino fluxes predicted by 
the Standard Solar Model by Bahcall and Pinsonneault (BP98)~\cite{BP98} 
taking into account the eccentricity of the Earth orbit around the Sun.

We do a $\chi^2$ analysis to fit these parameters 
and an extra normalization factor $f_B$ for the $^8$B neutrino 
flux, to the most recent experimental results coming from 
Homestake~\cite{homestake} $R_{\mbox{Cl}}= 2.56 \pm 0.21$ SNU, 
GALLEX\cite{gallex}  and SAGE\cite{sage} combined 
$R_{\mbox{Ga}}= 72.5 \pm 5.5$ SNU and SK~\cite{sk99} 
$R_{\mbox{SK}}= 0.475 \pm 0.015$ normalized to BP98. 
We use the same definition of the $\chi^2$ function to be minimized  
as in Ref.\ \cite{chi2}, 
except that our theoretical estimations were computed by convoluting 
the survival probability given in Eq.\ (\ref{prob})
with the absorption cross sections~\cite{bhp}, the  neutrino-electron 
elastic scattering cross section with radiative
corrections~\cite{xsec} and the solar neutrino flux corresponding to 
each reaction, $pp$, $pep$, $^7$Be, $^8$B, $^{13}$N and  $^{15}$O;
other minor sources were neglected.   

\begin{figure}
\centering\leavevmode
\epsfxsize=200pt
\epsfbox{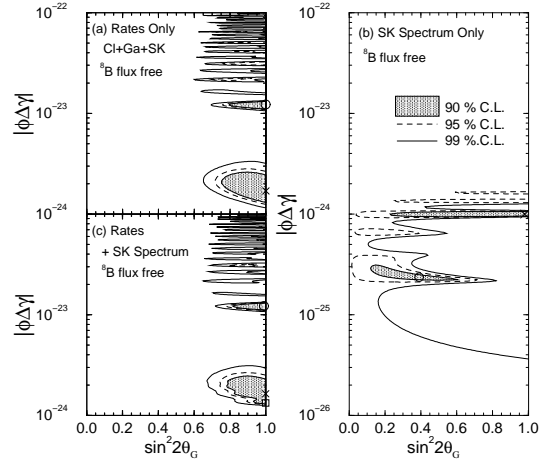}
\vglue -1.cm
\caption{
Allowed region for $\sin^22\theta_G$ and 
$|\phi \Delta \gamma|$ for (a) the rates only, (b) 
SK spectrum only and (c) rates + SK spectrum 
combined.  
The best fit points are indicated by the crosses
and the local best fit points in the other 90\% C.L. 
islands are indicated, in each plot, by the open circles. 
}
\label{fig1}
\vglue -0.35cm
\end{figure}

We present in Fig.\ 1 (a) the allowed region determined only by the
rates with free $f_B$, for fixed $^8$B flux ($f_B=1$) the allowed region is similar. In Ref.\ \cite{us} one can find a table which 
gives  more details on  best fitted parameters as well 
as the $\chi^2_{\mbox{\scriptsize min}}$ values for fixed and free $f_B$. 
We found for $f_B=1$ that $\chi^2_{\mbox{\scriptsize min}} = 1.49$ for 3-2=1 degree of freedom and for $f_B=0.81$ that $\chi^2_{\mbox{\scriptsize min}} = 0.32$ for 3-3=0 degree of freedom.

We then perform a spectral shape analysis fitting the $^8$B 
spectrum measured by SK ~\cite{sk99} using the following 
$\chi^2$ definition:
\begin{equation} 
\chi^2 = \sum_i 
\left[\frac{S^{\mbox{\scriptsize obs}}(E_i)-f_B S^{\mbox{\scriptsize theo}}(E_i)}{\sigma_i}\right]^2,
\end{equation}
where the sum is performed over all the 18 experimental points
$S^{\mbox{\scriptsize obs}}(E_i)$ normalized by BP98 prediction for 
the recoil-electron energy $E_i$, $\sigma_i$ is the total
experimental error and $S^{\mbox{\scriptsize theo}}$ is our theoretical 
prediction that was calculated using the  BP98 $^8$B differential flux, 
the $\nu-e$  scattering cross section~\cite{xsec}, the survival 
probability as given by Eq.\ (\ref{prob}) taking into account 
the eccentricity as we did for the rates, the experimental energy 
resolution as in Ref.\ \cite{res} and the detection efficiency as 
a step function with threshold $E_{\mbox{\scriptsize th}}$ = 5.5 MeV.

After the $\chi^2$ minimization with $f_B=0.80$ we have obtained 
$\chi^2_{\mbox{\scriptsize min}}=15.8$ for 18-3 =15 degree of freedom. 
The best fitted parameters that also can be found in Ref.~\cite{us} 
permit us to compute the allowed region displayed in Fig.\ 1 (b). 

Finally, we perform a combined fit of the rates and the spectrum 
obtaining the allowed region presented in Fig.\ 1 (c). 
The combined allowed region is essentially the same as the one 
obtained by the rates alone.  In all cases presented in 
Figs.\ 1 (a)-(c) we have two isolated islands of  90\% C.L. allowed 
regions. See Ref.\ \cite{us} for a table with the best fitted parameters 
for this global fit as well as for the fitted values corresponding to
the local minimum in these islands.  Note that only the upper 
corner of the Fig.\ 1 (c), for $|\phi \Delta \gamma | > 2 \times 10^{-23}$ 
and maximal mixing in the $\nu_e \to \nu_\mu$ channel can be excluded
by CCFR~\cite{pkm}. Moreover, there are no restrictions in the 
range of parameters we have considered in the case of 
$\nu_e \to \nu_\tau, \nu_s$ oscillations.

\section{DISCUSSIONS AND CONCLUSIONS}

Let's finally remark that, in contrast to the usual vacuum oscillation
solution to the SNP, in this VEP 
scenario no strong seasonal effect is expected in any of the present 
or future experiments, even the ones that will be sensitive to  
$^7$Be neutrinos~\cite{borexino,hellaz}. 
Contrary to the usual vacuum oscillation case, 
the oscillation length for the low energy
$pp$ and $^7$Be neutrinos are very large, comparable to or only a few times
smaller than the Sun-Earth distance, so that the effect of the 
eccentricity in the oscillation probability is small. 
On the other hand, for higher energy neutrinos relevant for 
SK, the effect of the eccentricity in the probability could be large, but 
it is averaged out after the integration over a certain neutrino energy range. 
These observations are confirmed by Fig.\ 4 of Ref.\ \cite{us}.


We have found a new solution to the SNP which is comparable in 
quality of the fit to the other suggested 
ones and can, in principle, be  discriminated from them in the near future.
In fact  a very-long-baseline neutrino experiment in a $\mu$-collider~\cite{geer} could directly probe the entire parameter region 
where this solution was found.

\section*{ACKNOWLEDGMENTS}

We thank P. Krastev, E. Lisi, G. Matsas, H.
Minakata, M. Smy, P. de Holanda and GEFAN for valuable discussions and comments.
H.N. thanks W. Haxton and B. Balantekin and the Institute for
Nuclear Theory  at the University of Washington for their hospitality 
and the Department of Energy for partial support during the 
final stage of this work. 
This work was supported by the Brazilian funding agencies FAPESP and CNPq.
%
%
\vspace{-0.1truecm}

\end{document}